\documentclass[11pt]{article}
\usepackage{ragged2e}
\usepackage{jheppub}
\usepackage{dsfont}
\usepackage{amsmath,amssymb,amscd,amsfonts,mathtools}
\usepackage{xcolor}
\definecolor{markgreen}{RGB}{230,243,230}
\definecolor{darkolivegreen}{rgb}{0.33, 0.42, 0.18}
\definecolor{darkpastelgreen}{rgb}{0.01, 0.75, 0.24}
\DeclareMathOperator{\Tr}{Tr}

\usepackage[toc,page]{appendix}
\usepackage{epsfig}
\usepackage{epstopdf}
\usepackage{latexsym}
\usepackage{graphicx}
\usepackage{booktabs}
\usepackage{bbm}
\usepackage{color}
\usepackage{physics}
\usepackage{tensor}
\usepackage{verbatim}
\usepackage{subfig}
\usepackage{tikz}
\usepackage{ifthen}
\usetikzlibrary{matrix}
\usetikzlibrary{decorations.markings,calc,shapes,decorations.pathmorphing}
\usetikzlibrary{patterns}
\usetikzlibrary{positioning}

\pdfoutput=1

\newboolean{showcomments}
\setboolean{showcomments}{false}
\newcommand\rem[1]{\ifthenelse{\boolean{showcomments}}{{#1}}{}}
\newcommand{\be}{\begin{equation}}
\newcommand{\ee}{\end{equation}}

\title{\Large BCFT in a Black Hole Background: An Analytical Holographic Model}
\author{Hao Geng, Lisa Randall and Erik Swanson}
\affiliation{Jefferson Physical Laboratory, Harvard University, 17 Oxford St., Cambridge, MA, 02138, USA.}

\emailAdd{haogeng@g.harvard.edu}
\emailAdd{randall@g.harvard.edu}
\emailAdd{eswanson@college.harvard.edu}

\abstract{We study the entanglement phase structure of a holographic boundary conformal field theory (BCFT) in a two-dimensional black hole background. The bulk dual is the AdS$_3$ black string geometry with a Karch-Randall brane. We compute the subregion entanglement entropy of various two-sided bipartitions to elucidate the phase space where a Page curve exists in this setup. We do fully analytical computations on both the gravity side and the field theory side and  demonstrate that the results precisely match. We discuss the entanglement phase structure describing where a Page curve exists in this geometry in the context of these analytical results. This is a useful model to study entanglement entropy for quantum field theory on a curved background.}

\begin{document}	
\maketitle
\flushbottom
\section{Introduction}
The Karch-Randall braneworld \cite{Karch:2000ct,Karch:2000gx} has taught us important lessons about quantum gravity. It allows for constructing calculable models of entanglement islands \cite{Almheiri:2019yqk,Almheiri:2019psy,Geng:2020qvw,Chen:2020uac} with the potential to partially resolve a version of the black hole information paradox. This involves demonstrating the existence of a unitary Page curve for black hole radiation \cite{Almheiri:2019psy,Geng:2020qvw,Chen:2020hmv} that in some cases can be calculated analytically  \cite{Geng:2020qvw}.
In higher-dimensional setups, these advances have exploited the three equivalent descriptions of the Karch-Randall braneworld:
\begin{itemize}
    \item \textbf{The bulk description:} A d-dimensional end of the world brane embedded in an asymptotically AdS$_{d+1}$ space. The geometry of the brane is asymptotically AdS$_{d}$ and the physics of the AdS$_{d+1}$ bulk is described by pure Einstein's gravity.
    
    \item \textbf{The intermediate description:} A quantum gravity theory on an asymptotically AdS$_d$ space glued to a half space conformal field theory by imposing transparent boundary conditions on their common boundary. 
    
    \item \textbf{The boundary description:} A d-dimensional conformal field theory living on a manifold with boundary where we impose conformal boundary conditions. This is also called a boundary conformal field theory (BCFT).
    
\end{itemize}
The intermediate picture provides the context in which we can study the information transfer between a d-dimensional black hole and a thermal bath. The bath is realized by the half space conformal field theory, which absorbs the radiation from the black hole. The entanglement entropy of the radiation can then be formulated as the entanglement entropy of a specific bipartition in the boundary description. In a generic case the calculation of the entanglement entropy for a boundary subregion $\mathcal{A}$ is done in the bulk description using the Ryu-Takayanagi formula \cite{Ryu:2006bv,Fujita:2011fp}
\begin{equation}
    S_{\mathcal{A}}=\frac{\text{Area}(\gamma)}{4G_{N}}\,,
\end{equation}
where $\gamma$ is the Ryu-Takayanagi surface, which is a bulk minimal surface homologous to the boundary subregion $\mathcal{A}$, and $G_{N}$ is the bulk Newton's constant.

 The authors of  \cite{Geng:2021mic} successfully studied interesting aspects of conformal field theory on a specific curved space: the four-dimensional eternal AdS Schwarzschild black hole with a conformal boundary condition imposed on its asymptotic boundary. We emphasize that in this context the black hole is not gravitating but it is still radiating.\footnote{The radiating process equilibrates due to the conformal boundary condition for the stress-energy tensor $T_{\perp\parallel}=0$ imposed on its asymptotic boundary.} For simplicity that paper  studied  only the entanglement entropy of symmetric  bipartitions of the system. As we will review in Sec.~\ref{sec:review}, these bipartitions are two-sided, and capture some interesting dynamical aspects of the system. The main result is an interesting phase structure of the entanglement entropy of the bipartition; i.e. for certain bipartitions we would see time-dependent entanglement entropy and for others the entanglement entropy would be constant. The calculation in \cite{Geng:2021mic} is done for a four-dimensional AdS Schwarzschild black hole (with the dual bulk geometry as five-dimensional black string in AdS). It is fully numerical and  relies on applying the Ryu-Takayanagi formula back in the bulk description. 
 
 %The information paradox regarding the entanglement entropy of the black hole radiation can be made more precise using the boundary description.
However, in most entanglement entropy calculations such as the one described above, the boundary description in terms of the field theory is not fully exploited. 
In this paper, we  provide a model in a lower-dimensional context in which the calculations can be done in both  the boundary description and the bulk description fully analytically, even with more general asymmetric bipartitions. We first do the computation in the bulk description using the Ryu-Takayanagi formula i.e. the gravity side calculation. Then we perform the calculation in the boundary field theory description and find that it  indeed matches the gravity side calculation. This is a nontrivial check of the equivalence between the bulk description and the boundary description which is also called the AdS/BCFT correspondence \cite{Fujita:2011fp,Takayanagi:2011zk,Sully:2020pza}. We then provide a detailed analysis of the entanglement phase structure using our analytical results and discuss them in the concluding sections.

% Moreover, motivated by the fact that the gravity localized on the Karch-Randall braneworld is a massive gravity theory it has been conjectured in \cite{Geng:2020qvw}, further exploited in \cite{Geng:2020fxl} and later proved \cite{Geng:2021hlu} that entanglement islands exist only in massive gravity theories for a large class of spacetimes including asymptotically anti-de Sitter (AdS) spaces.\footnote{See \cite{Demulder:2022aij} for a recent string theory reaalization. 

\section{Review of Previous Work}\label{sec:review}
\subsection{Holographic BCFT in a Black Hole Background}
The paper \cite{Geng:2021mic} considered the black string geometry in AdS$_{d+1}$, which has the following metric
\begin{equation}
    ds^2=\frac{1}{u^2\sin^{2}\mu}\Bigg[-\left(1-\frac{u^{d-1}}{u_{H}^{d-1}}\right)dt^2+\frac{du^2}{1-\frac{u^{d-1}}{u_{H}^{d-1}}}+d\vec{x}^2_{d-2}+u^2d\mu^2\Bigg]\,,\label{eq:metricd}
\end{equation}
where $\mu\in[0,\pi]$ and $\mu=0\cup \mu=\pi$ is the asymptotic boundary. 
A Karch-Randall brane  is embedded in the bulk geometry as a constant-$\mu$ slice. The brane cuts off the bulk region behind it and if the brane sits at $\mu=\mu_{B}$ then the leftover bulk region runs from $\mu=\mu_{B}$ to $\mu=\pi$ (see Fig.~\ref{fig:LUP}). Using the AdS/BCFT correspondence, the field theory dual of this geometry at a generic value of $d$ is a boundary conformal field theory living on an AdS$_d$ black hole background with conformal boundary condition imposed at its asymptotic infinity.

\begin{figure}
    \centering
    \begin{tikzpicture}
       \draw[-,very thick] 
       decorate[decoration={zigzag,pre=lineto,pre length=5pt,post=lineto,post length=5pt}] {(-2.5,0) to (2.5,0)};
       \draw[-,very thick,red] (-2.5,0) to (-2.5,-5);
       \draw[-,very thick,red] (2.5,0) to (2.5,-5);
         \draw[-,very thick] 
       decorate[decoration={zigzag,pre=lineto,pre length=5pt,post=lineto,post length=5pt}] {(-2.5,-5) to (2.5,-5)};
       \draw[-,very thick] (-2.5,0) to (2.5,-5);
       \draw[-,very thick] (2.5,0) to (-2.5,-5);
       \draw[->,very thick,black] (-2.7,-3.5) to (-2.7,-1.5);
       \node at (-3,-2.5)
       {\textcolor{black}{$t_{L}$}};
        \draw[->,very thick,black] (2.7,-3.5) to (2.7,-1.5);
       \node at (3,-2.5)
       {\textcolor{black}{$t_{R}$}};
       \draw[-,thick, green] (-2.5,-2.5) to (-1.5,-2.5);
       \draw[-,thick, green] (2.5,-2.5) to (1.5,-2.5);
       \draw[-,thick, blue] (-1.5,-2.5) to (1.5,-2.5);
         \node at (-1.5,-2.2)
         {\textcolor{black}{$u_{L}$}};
          \node at (1.5,-2.2)
         {\textcolor{black}{$u_{R}$}};
         \node at (-1.5,-2.5)
         {\textcolor{black}{$\bullet$}};
        \node at (1.5,-2.5)
         {\textcolor{black}{$\bullet$}};
         %%%
          \draw[-,very thick] 
       decorate[decoration={zigzag,pre=lineto,pre length=5pt,post=lineto,post length=5pt}] {(2.7+1.4,0) to (7.7+1.4,0)};
       \draw[-,very thick,red] (2.7+1.4,0) to (2.7+1.4,-5);
       \draw[-,very thick,red] (7.7+1.4,0) to (7.7+1.4,-5);
         \draw[-,very thick] 
       decorate[decoration={zigzag,pre=lineto,pre length=5pt,post=lineto,post length=5pt}] {(2.7+1.4,-5) to (7.7+1.4,-5)};
       \draw[-,very thick] (2.7+1.4,0) to (7.7+1.4,-5);
       \draw[-,very thick] (7.7+1.4,0) to (2.7+1.4,-5);
       \draw[->,very thick,black] (2.5+1.4,-3.5) to (2.5+1.4,-1.5);
       \node at (2.2+1.4,-2.5)
       {\textcolor{black}{$t_{L}$}};
        \draw[->,very thick,black] (7.9+1.4,-3.5) to (7.9+1.4,-1.5);
       \node at (8.2+1.4,-2.5)
       {\textcolor{black}{$t_{R}$}};
       \draw[-,thick, green] (2.7+1.4,-2.5+1) to (3.7+1.4,-2.5+0.6);
       \draw[-,thick, green] (7.7+1.4,-2.5+1) to (6.7+1.4,-2.5+0.6);
       \draw[-,thick, blue] (3.7+1.4,-2.5+0.6) to
       (5.2+1.4,-2.5);
       \draw[-,thick, blue] (5.2+1.4,-2.5) to (6.7+1.4,-2.5+0.6);
         \node at (3.7+1.4,-2.2)
         {\textcolor{black}{$u_{L}$}};
          \node at (6.7+1.4,-2.2)
         {\textcolor{black}{$u_{R}$}};
         \node at (3.7+1.4,-2.5+0.6)
         {\textcolor{black}{$\bullet$}};
        \node at (6.7+1.4,-2.5+0.6)
         {\textcolor{black}{$\bullet$}};
    \end{tikzpicture}
    \caption{Two diagrams representing different time slices on the boundary ($\rho=\infty$) Penrose diagram of the eternal black hole on which the BCFT$_d$ lives. The two red vertical lines represent the asymptotic boundaries where we impose conformal boundary conditions. Each time slice is composed of a blue component on one side of the bipartition and a green component on the other side. The green intervals, determined by $u_L$ and $u_R$ are drawn as equal size here, but generically vary in our setup. The union of the two green intervals is the subsystem we considered in \cite{Geng:2021mic} where we computed the entanglement entropy between the green and the blue subsystems. We label time as $t_L$ and $t_R$ on either side of the diagram, and for both sides we take time evolution to go up the diagram (note the contrast with Figure \ref{fig:Penrose1}). The $u$-coordinate increases along a given time slice from 0 on the red boundary to $u_H$ at the bifurcation horizon, where the black diagonal lines cross. At the jagged singularities, $u=\infty$.}
    \label{pic:vanishing}
\end{figure}
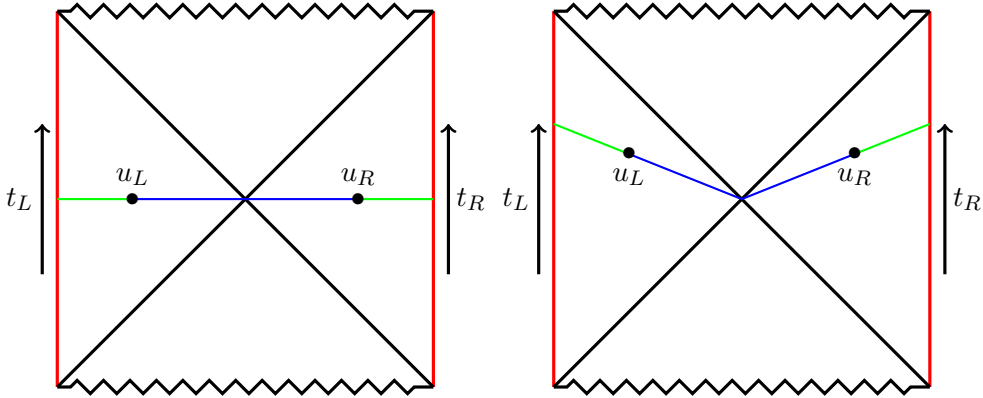

More precisely, the geometry on each constant$-\mu$ slice of the bulk black string is an eternal AdS$_d$ black hole. Hence the dual BCFT$_d$ is living on an eternal black hole background which has two asymptotic boundaries (see Fig.\ref{pic:vanishing} for the boundary Penrose diagram). The paper \cite{Geng:2021mic}  considered the  the entanglement entropy of the field theory subsystem as indicated by the green intervals and time evolved the system as in Fig.~\ref{pic:vanishing} to study whether or not there is a time-dependent entanglement entropy i.e. a nontrivial Page curve. The boundaries (away from the asymptotic boundary) of the green intervals are denoted as $u_{L}$ and $u_{R}$ (in the coordinate defined by Equ.~(\ref{eq:metricd})) and  \cite{Geng:2021mic}  considered only the case $u_{L}=u_{R}$. 

To address this question it suffices to  look at the zero-time slice using holography. The reason is that there are two candidate Ryu-Takayanagi (RT) surfaces for the entanglement entropy, with one going through the black string interior connecting the boundaries of the two green intervals and the other staying outside the black string horizon connecting the boundaries of the green intervals to the nearby branes, and the RT prescription tells us to take the area of the smaller one to compute the entanglement entropy. The one that goes through the black string interior, which is called the Hartman-Maldacena surface,  will monotonically grow with time (as pointed out by Hartman and Maldacena in \cite{Hartman:2013qma}) and the one that stays outside the black string interior, the island surface, will be constant in time. The island surface has two disconnected components with one in each exterior region of the bulk black string. Therefore, if the Hartman-Maldacena surface has the smaller area at zero time then we will have a time-dependent entanglement entropy where initially the entanglement entropy is calculated by the Hartman-Maldacena surface and later switches to the island surface when the area of the Hartman-Maldacena surface grows beyond that of the island surface. Otherwise, there is a constant entanglement entropy which is computed by the island surface. As it was firstly noticed in \cite{Geng:2020qvw} and later pointed out in \cite{Geng:2020fxl}, the later case with a constant entanglement entropy can be understood as the black hole being a fast scrambler.  In this case the Hilbert space of the subsystem we are considering is presumably too small to be further scrambled. By considering the simple case $u_{L}=u_{R}=u_{0}$ and  doing the calculation only on the zero-time slice  Ref.\cite{Geng:2021mic} mapped out a phase diagram parameterized by $u_{0}$ and the brane angle $\mu_{B}$ for parameter regions where we can and cannot have a time-dependent entanglement entropy. The calculation in \cite{Geng:2021mic} was fully numerical for $d=5$.

In this paper, we  consider the lower dimensional case $d=2$ which is tractable even for  $u_{L}\neq u_{R}$ and  beyond the zero-time slice. The calculation in this paper is fully analytical and can be done on both the gravitational side using holography and the dual field theory side with mutually matched results. We  map out a more complete phase diagram for this lower-dimensional case as compared with \cite{Geng:2021mic}.

\subsection{Calculations of Entanglement Entropy in 2d Holographic BCFT}\label{sec:reviewCFT}

In this section, we review the calculation of entanglement entropy in two-dimensional boundary conformal field theories (BCFT$_2$'s) and its simplification for BCFTs with holographic duals. For convenience, we will closely follow \cite{Sully:2020pza,Geng:2021iyq} to use BCFT$_2$ on a flat background to demonstrate the concepts and techniques for the calculation in this section. In later sections, we will apply these techniques to our study of BCFT$_2$ in an AdS black hole background.
\subsubsection{Vacuum State}

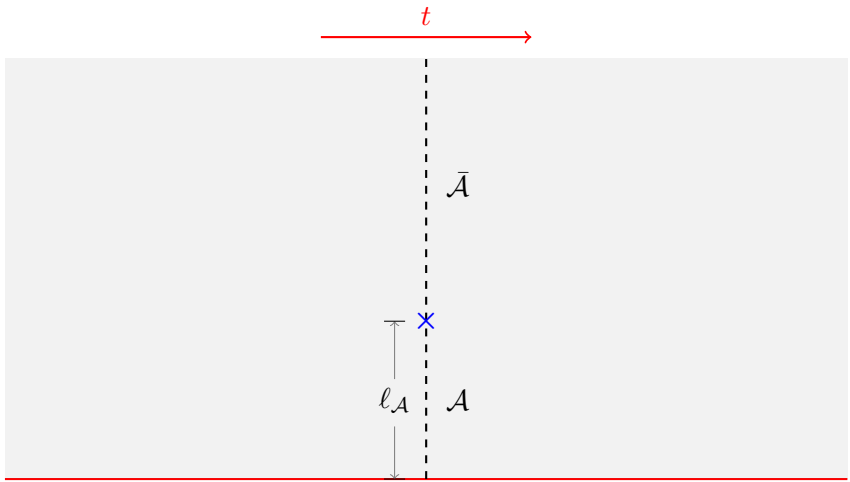
\begin{figure}[h]
\begin{centering}
{
\begin{tikzpicture}[scale=1.4]
\draw[-, thick,red] (-4,0) to (4,0);

%\draw[pattern=north west lines,pattern color=gray!200,draw=none] (-4,0) to (4,0) to (4,4) to (-4,4);
\draw[fill=gray, draw=none, fill opacity = 0.1] (-4,0) to (4,0) to (4,4) to (-4,4) to (-4,0);
\draw[->, thick,red] (-1,4.2) to (1,4.2);
\node at (0,4.4)
{\textcolor{red}{$t$}};
%\draw[-,dashed,color=black!50!lime,very thick] (0,0) to (0,4);
\draw[-,dashed,color=black, thick] (0,0) to (0,4);
\node at (0,1.5) {\textcolor{blue}{$\cross$}};
\node at (0.3,0.75)
{\textcolor{black}{$\mathcal{A}$}};
\node at (0.3,2.8)
{\textcolor{black}{$\bar{\mathcal{A}}$}};
\node at (-0.3,0.75)
{\textcolor{black}{$\ell_\mathcal{A}$}};
\draw[->, thin,gray] (-0.3,0.95) to (-0.3,1.5);
\draw[->, thin,gray] (-0.3,0.5) to (-0.3,0);
\draw[-, thin,black] (-0.4,1.5) to (-0.2,1.5);
\draw[-, thin,black] (-0.4,0) to (-0.2,0);
\end{tikzpicture}
}
\caption{\small{\textit{This diagram shows the situation for a BCFT$_2$ living on the upper half plane (UHP). We consider bulk CFT to be in the vacuum state. The boundary is specified by the red horizontal axis where we impose conformal boundary conditions. The time direction is along the horizontal axis. We take a constant time slice (dashed black vertical line) that defines the quantum state  we are studying. Our goal is to compute the entanglement entropy associated with the bipartition indicated by the blue cross.}}}
\label{demonEE}
\end{centering}
\end{figure}

To review the calculation of entanglement entropy in 2d BCFTs, let's consider the case that the bulk CFT$_2$ is in the vacuum state. For the sake of convenience, we will consider the Euclidean signature. In this case, the CFT is living on an upper half plane parameterized by the complex coordinates $(z,\bar{z})$ with $\Re{z}\geq0$. The conformal boundary condition is imposed on the boundary $\Re{z}=0$ and it preserves half of the conformal invariance of bulk CFT$_2$ \cite{Cardy:2004hm}. The Euclidean time $t$ is going along the axis of $\Re{z}$ (see Fig.~\ref{demonEE}). We will compute the entanglement entropy associated with the bipartition (the blue cross in Fig.~\ref{demonEE}) indicated in Fig.~\ref{demonEE}. This bipartition factorizes the whole system into $\mathcal{A}$ and its complement $\bar{\mathcal{A}}$ and we primarily focus on $\mathcal{A}$ and use its reduced density matrix $\rho_{\mathcal{A}}$ to compute this entanglement entropy. This entanglement entropy is computed by taking the limit $n\rightarrow1$ of the $n$-th Renyi entropy
\begin{equation}
    S_{\mathcal{A}}^{n}=-\frac{1}{n-1}\ln\Tr(\rho_{\mathcal{A}}^n) \,. \label{eq:defintion}
\end{equation}
To compute the trace $\Tr(\rho_{\mathcal{A}}^n)$ in the above formula we can use the replica trick \cite{Calabrese:2009qy}. The result is that this is equivalent to compute the one-point function of a twist operator $\Phi_{n}(z,\bar{z})$ inserted at the bipartition point (the blue cross in Fig.~\ref{demonEE}). The effect of the twist operator is creating a branch cut on the upper half plane (UHP) which is equivalent to considering a smooth 2d manifold with multiple covers as obtained from the replica trick. The branch cut is from the blue cross to infinity along the dashed black line in Fig.~\ref{demonEE}.

As a result, the entanglement entropy of the bipartition that we are considering is translated to the following formula
\begin{equation}
    S_{\mathcal{A}}=\lim_{n\rightarrow1}\frac{1}{1-n}\ln\langle\Phi_{n}(z,\bar{z})\rangle_{\text{UHP}} \,, \label{eq:von}
\end{equation}
where $(z,\bar{z})$ is the coordinate of the bipartition point on the UHP.

It has been understood that the twist operator $\Phi_{n}(z,\bar{z})$ is a primary operator \cite{Calabrese:2009qy} with conformal dimensions
\begin{equation}
    h_{n}=\bar{h}_{n}=\frac{c}{24}\left(n-\frac{1}{n}\right).
\end{equation}
Hence the one-point function in Equ.~(\ref{eq:von}) can be fixed by conformal symmetry. The result is similar in the computation of the Green's function in electrodynamics when there is a dielectric boundary. It can be computed by the doubling trick \cite{Sully} the UHP $\langle\Phi_{n}(z,\bar{z})\rangle_{\text{UHP}}$ is equal to a two-point function for a chiral primary field $\Phi_{n}(z)$ (with conformal dimensions $h_n=\frac{c}{24}\left(n-\frac{1}{n}\right),\,\bar{h}_{n}=0$) on the complex plane. More precisely, this two point function is for one $\Phi_n$ inserted at the bipartition point and the other $\Phi_n$ at its mirror symmetric point on the lower half plane,
\begin{equation}
        \langle\Phi_{n}(z,\bar{z})\rangle_{\text{UHP}}=\langle\Phi_{n}(z)\Phi_{n}(z^*)\rangle_{\mathbb{C}}=\frac{\mathcal{A}_{\Phi_{n}}^{b}}{|z-z^{*}|^{2h_{n}}} \,,\label{eq:1pUHP}
\end{equation}
where $\mathcal{A}_{\Phi_{n}}^{b}$ is a normalization constant that is determined by the specific conformal boundary condition for the BCFT$_2$ \cite{Sully:2020pza} and will be fixed below.

Another way to compute the one-point function $\langle\Phi_{n}(z,\bar{z})\rangle_{\text{UHP}}$ is to do the boundary operator expansion (BOE) for the bulk operator $\Phi_{n}(z\bar{z})$ first and compute the expectation value of the resulting operator sum. The BOE states that for a BCFT on the UHP any bulk operator $\mathcal{O}_{i}(z,\bar{z})$ can be expanded as sum of boundary operators $\hat{O}_{I}(x)$,
\begin{equation}
    \mathcal{O}_{i}(z,\bar{z})=\sum_{J}\frac{\mathcal{B}_{i}^{bJ}}{(2y)^{\Delta_{i}-\Delta_{J}}}\tilde{\mathcal{C}}[y,\partial_{x}]\hat{\mathcal{O}}_{J}(x) \,.\label{eq:BOE}
\end{equation}
In this expression $\Delta_{i}$ is the conformal weight of $\mathcal{O}_{i}(z,\bar{z})$ (i.e. $h_{i}=\bar{h}_{i}=\frac{\Delta_{i}}{2}$), $\Delta_{J}$ is that of the boundary primary operator $\hat{\mathcal{O}}_{J}(x)$ and we use the complex coordinate $z=x+iy$ ($y>0$) on the UHP. The coefficients $\mathcal{B}_{i}^{bJ}$ are the so called the BOE coefficients and they are determined by the boundary condition and the structure of the parent CFT. The boundary primary operator is normalized as
\begin{equation}
    \langle\hat{O}_{I}(x_{I})\hat{O}_{J}(x_{J})\rangle=\frac{G_{IJ}}{|x_{I}-x_{J}|^{2\Delta_{I}}} \,.
\end{equation}
where we have $\mathcal{B}_{iI}^{b}=\sum_{J}\mathcal{B}_{i}^{bJ}G_{IJ}$. Matching the result from the doubling trick Eq.~\eqref{eq:1pUHP} with the vacuum expectation value of the BOE expansion Eq.~\eqref{eq:BOE}, we see that the only boundary operator in the BOE that contributes to $\left<\Phi_n\right>_{\text{UHP}}$ is the identity operator $\mathbf{1}$, and we have 
\begin{equation}
    \mathcal{A}_{\Phi_{n}}^{b}=\mathcal{B}_{\Phi_{n}\mathbf{1}}^{b} \,.
\end{equation}

As result we have the following explicit expression for entanglement entropy of the bipartition we are considering:
\begin{equation}
    S_{\mathcal{A}}=\frac{c}{6}\ln(\frac{2\ell_{\mathcal{A}}}{\epsilon})+\ln(g_{b}) \,,\label{eq:EEvac}
\end{equation}
where $\ell_{\mathcal{A}} = (z-\bar{z})/2i$ is the length of the subsystem $\mathcal{A}$ and the second term is called the boundary entropy. It is defined through the regularization 
\begin{equation}
    \ln(g_{b})-\frac{c}{6}\ln(\epsilon)=\lim_{n\rightarrow1}\frac{1}{1-n}\ln(B_{\Phi_{n}1}^{b})\, ,
    \label{eq:bdyentropy}
\end{equation}
where $\epsilon$ is a UV cutoff that universally appears in the entanglement entropy of quantum field theories. We emphasize that the result Equ.~(\ref{eq:EEvac}) matches the holographic computation in \cite{Takayanagi:2011zk} and this is the universal form for the bipartite entanglement entropy of the vacuum state in any BCFT on the UHP when the background geometry is flat.

\subsubsection{Thermal field double state}\label{sec:ReviewTFD}
Equipped with the concepts and techniques reviewed in the former section, we will consider a more relevant case to black holes in this section. We consider the thermal field double (TFD) state 
of two 2d BCFTs. We emphasize again that the two BCFTs in this section are all in flat background.
We will label the two BCFTs separately as L and R.

\begin{figure}[h]
\begin{centering}
\subfloat[\textit{The thermofield double state and its time evolution}\label{TFDpre1}]
{
\begin{tikzpicture}[scale=0.9]
\draw[-,color=red, thick] (-1,0) arc (1:361:-1);
\draw[-,dashed,color=black, thick] (-3.5,0) to (-1,0);
\draw[-,dashed,color=black, thick] (1,0) to (3.5,0);
\node at (-2,-0.4)
{\textcolor{black}{BCFT$_L$}};
\node at (2,-0.4)
{\textcolor{black}{BCFT$_R$}};
%\draw[-,dashed,color=darkpastelgreen, thick](0.705,0.705) to (1.41+0.705,1.41+0.705);
%\draw[-,dashed,color=darkpastelgreen, thick](-0.705,0.705) to (-1.41-0.705,1.41+0.705);
\draw[-,dashed,color=darkpastelgreen, thick](0.705+0.5,0.705+0.5) to (1.41+0.705,1.41+0.705);
\draw[-,dashed,color=darkpastelgreen, thick](-0.705-0.5,0.705+0.5) to (-1.41-0.705,1.41+0.705);
\draw[-,color=darkpastelgreen, thick](0.705,0.705) to (0.705+0.5,0.705+0.5);
\draw[-,color=darkpastelgreen, thick](-0.705,0.705) to (-0.705-0.5,0.705+0.5);

\draw[<-,, color=red, thin] (0.7,0.4) arc (25:-25:1);
\node at (0.45,0.2) {\textcolor{black}{$t_R$}};

\draw[<-,, color=red, thin] (-0.7,0.4) arc (-25:25:-1);
\node at (-0.45,0.2) {\textcolor{black}{$t_L$}};

\node at (0.705+0.5,0.705+0.5) {\textcolor{blue}{$+$}};
\node at (-0.705-0.5,0.705+0.5) {\textcolor{blue}{$+$}};
\node at (0.705,0.705+0.5) {\textcolor{black}{$\mathcal{A}_R$}};
\node at (-0.705,0.705+0.5) {\textcolor{black}{$\mathcal{A}_L$}};
\node at (0.705+0.75,0.705+0.5+0.75) {\textcolor{black}{$\bar{\mathcal{A}}_R$}};
\node at (-0.705-0.75,0.705+0.5+0.75) {\textcolor{black}{$\bar{\mathcal{A}}_L$}};
\end{tikzpicture}
}
\hspace{0.5cm}
\subfloat[\textit{Conformal mapping of the region on left to a UHP} \label{TFDmap1}]
{
\begin{tikzpicture}[scale=0.9]

\draw[fill=gray, draw=none, fill opacity = 0.1] (-4,0) to (4,0) to (4,3) to (-4,3);
\draw[-, thick,red] (-4,0) to (4,0);

\node at (1.5,1.5) {\textcolor{blue}{$\cross$}};
\node at (-1.5,1.5) {\textcolor{blue}{$\cross$}};
\draw[-,dashed,darkpastelgreen] (-1.5,1.5) to (1.5,1.5);
\node at (-1.5,1.9) {\textcolor{black}{\emph{$\bar{\Phi}_{n}$}}};
\node at (1.5,1.9) {\textcolor{black}{\emph{$\Phi_{n}$}}};
\node at (0,2) {\textcolor{black}{\emph{$\bullet$}}};
\node at (0,2.4) {\textcolor{black}{\emph{$\left(0,\frac{1}{2}\right)$}}};
\end{tikzpicture}

}
\caption{\small{\textit{a) The two Euclidean BCFTs L and R in the TFD state: time evolution is rotation with respect to the origin, hence the red circle is the time evolution of the boundary. The two black dashed lines define the zero time slice.
Under the chosen time evolution, L and R evolve clockwise and counter-clockwise respectively, as indicated.
We are interested in the entanglement entropy of the subsystem $\mathcal{A}=\mathcal{A}_{L}\cup\mathcal{A}_{R}$ corresponding to the solid green line segments. 
b) The UHP which results from the conformal mapping of the region outside of the red circle in Fig.~\ref{TFDpre1}. The circular boundary is mapped to the real axis and infinity is mapped to $\left(0,\frac{1}{2}\right)$. The location of twist operators is mapped to the two blue crosses, separated in the horizontal direction. The branch cut is mapped (and deformed) to the dashed green line connecting the two operators.}}}
\label{TFDall}
\end{centering}
\end{figure}
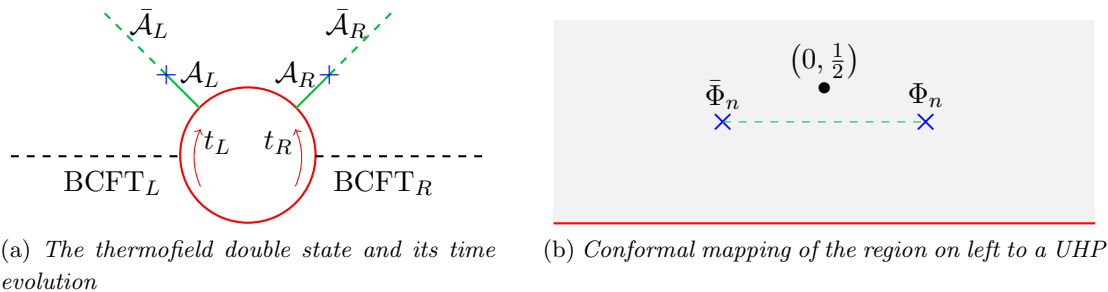

The TFD state can be prepared using an Euclidean path integral.
In this Euclidean picture, the time direction is periodic, and we can choose the time evolution for the L and R BCFTs such that the TFD state evolves non-trivially (see Fig.~\ref{TFDpre1}).  We would like to calculate the entanglement entropy of the bipartition shown in Fig.~\ref{TFDpre1} which factorizes the whole system into the subsystem $\mathcal{A}_L \cup \mathcal{A}_R$ and its complement. In analogy with the previous section, computing the entanglement entropy using the replica trick is equivalent to the computation of the two-point function of a twist operator $\Phi_n$ and an anti-twist operator $\bar{\Phi}_n$ inserted respectively at the two blue crosses in Fig.~\ref{TFDpre1}. Denoting the complex coordinate of the two crosses respectively by $w_L$ and $w_R$, the resulting two-point function is
\begin{equation}
    \langle\Phi_{n}(w_{R},\bar{w}_{R})\bar{\Phi}_{n}(w_{L},\bar{w}_{L})\rangle \,.
\end{equation}
In order to compute this two-point function, we will map the configuration to the upper half plane, which is parametrized by z coordinates, using the following conformal transformation
\begin{equation}
    w=\frac{1}{z-\frac{i}{2}}-i\label{eq:TFDmap1} \,.
\end{equation}
Under this transformation, the boundary circle in $w$ coordinates (red circle in Fig.~\ref{TFDpre1}) is mapped to the real axis in the UHP and the infinity in $w$ is mapped to a point $z = i/2$. 
Moreover, the insertion points of the twist and anti-twist operators are mapped to two points on the UHP as shown in Fig.~\ref{TFDmap1}.

The task now is to compute the following two-point function of a twist operator and an anti-twist operator on the UHP:
\begin{equation}
    \langle\bar{\Phi}_{n}(z_{L},\bar{z}_{L})\Phi_{n}(z_{R},\bar{z}_{R})\rangle_{\text{UHP}} \,.\label{eq:2pt}
\end{equation}

In the absence of the boundary, this is the standard two-point function of primary operators in 2d CFT and it is totally fixed by the conformal symmetry as:
\begin{equation}
    \langle\bar{\Phi}_{n}(z_{L},\bar{z}_{L})\Phi_{n}(z_{R},\bar{z}_{R})\rangle_{\mathbb{C}}=\frac{\epsilon^{2d_{n}}}{|z_{L}-z_{R}|^{2d_{n}}} \,,\label{eq:bulkchannel}
\end{equation}
where $d_{n}$ denotes the conformal weight and it equals to $2h_{n}=2\bar{h}_{n}=\frac{c}{12}(n-\frac{1}{n})$. In this expression we have set the normalization of the two-point function to be $\epsilon^{2d_{n}}$, (where $\epsilon$ is a UV cutoff) and as we will show later that this is in order to have clean form for the entanglement entropy.

In this presence of the boundary, this two-point function is not easy to compute for a general 2d BCFT. The reason is that the doubling trick will translate it into a four-point function in a chiral CFT on the complex plane. In general, there is no universal form of the four point function in 2d CFT and it is determined by the details of the CFT. If we want to use the operator production expansion to compute this two-point functions there are two ways to do this. The first way is based on the observation that away from the boundary the structure of the BCFT is the same as its parent CFT. This tells us that the operator product expansion (OPE) for two bulk operators is the same as in the parent CFT:
\begin{equation}
    \mathcal{O}_{i}(z_{1},\bar{z}_{1})\mathcal{O}_{j}(z_{2},\bar{z}_{2})=\sum_{k}\frac{\hat{\mathcal{C}}_{ij}^{k}}{|z_{1}-z_{2}|^{\Delta_{i}+\Delta_{j}-\Delta_{k}}}C_{\Delta_{i}\Delta_{j}\Delta_{k}}[z_{12},\bar{z}_{12},\partial_{2},\bar{\partial}_{2}]\mathcal{O}_{k}(z_{2},\bar{z}_{2}) \,,\label{eq:OPE}
\end{equation}
for bulk primaries $\mathcal{O}_{k}$ 
and OPE coefficients $\mathcal{\hat{C}}_{ij}^{k}$. 
Hence to compute the two-point function Equ.~(\ref{eq:2pt}), we have to compute the expectation value of the resulting sum of bulk primary operators. This turns to be a sum over one-point functions of the bulk primary operators $\mathcal{O}_k$ appearing in the product, as per Eq.~\eqref{eq:1pUHP} and Eq.~\eqref{eq:BOE}. This is the so called \textit{bulk channel}.
Alternatively, we can perform the boundary operator expansion (BOE) as in Eq.~\eqref{eq:BOE} first and then compute the resulting (normalized) boundary two-point functions. This is called the \textit{boundary channel}.

The consistency of the BCFT tells us that these two channels should produce the same result. This is a nontrivial bootstrap constraint.
For holographic BCFTs, which is the case we will be interested in, this constraint will significantly simplify the computation. A holographic CFT is a CFT with holographic dual as Einstein's gravity in one-dimensional higher anti-de Sitter space \cite{Hartman:2013mia}. In the 2d CFT case, it is characterized by the following two properties \cite{Hartman:2013mia}:
\begin{itemize}
    \item The central charge $c$ is large.
    \item The spectrum of light operators, of conformal dimension $\mathcal{O}(c^{0})$,\footnote{In this paper we follow \cite{Geng:2021iyq} where the symbol $\mathcal{O}(\dots)$ is used to denote "of the order $\dots$".} is sparse.
\end{itemize}
The first condition implies that in the OPE the contribution from operators of conformal dimension $\mathcal{O}(c)$ or higher, which is usually referred to as \textit{heavy} operators, are suppressed (see \cite{Sully:2020pza,Hartman:2013mia,Faulkner:2013ana} for details). The second condition ensures the total contributions to the OPE from the light operators is a multiplication of the contribution of the identity operator. As a result, to a precise level of approximation, the whole OPE is proportional to the contribution from the identity operator and this called the \textit{vacuum dominance}. This greatly simplifies the calculation of the higher-point correlators \cite{Sully:2020pza,Hartman:2013mia,Faulkner:2013ana}.

However, in our case we consider holographic BCFTs and in this case the two requirements above are naturally generalized to boundary operators \cite{Sully:2020pza} though with some subtleties. Let us consider the bulk channel first. In the appearance of the boundary, the one-point functions of bulk operators are now generically nonzero, and in the calculation of the two-point function Equ.~(\ref{eq:2pt}), this effect may compete with the effect that contribution from the heavy operators are suppressed.
Nevertheless, this could happen only if we are close enough to the boundary such that the one-point functions of heavy operators are large enough. With this in mind, we should use the boundary channel to compute the two-point function Equ.~(\ref{eq:2pt}) when the two bulk operators are close enough to the boundary.  In this case, as it is shown in Eq.~\eqref{eq:BOE}, the contribution of a boundary operator with conformal dimension $\Delta_J$ to the BOE of $\Phi$ scales as $y^{\Delta_J}$. Hence for small enough $y$ the contribution from heavy boundary operators is suppressed and by the sparseness requirement we just have to look at the contribution from the identity operator. This is called the vacuum dominance in the boundary channel. However, for large $y$, the contribution from heavy boundary operators is not suppressed anymore. This time we should work in the bulk channel.

In summary, the use of the bulk channel vacuum dominance for the calculation of the bulk two-point function for holographic BCFTs works only when the operator insertions are far away from the boundary, and the vacuum dominance in the boundary channel applies only when close to the boundary. Consistency of the BCFT input a bootstrap constraint that the results calculated by the bulk channel and the boundary channel must match such that there is a unique result. Hence, we have to figure out the criterion for the applications of the bulk channel vacuum dominance and the boundary channel vacuum dominance to ensure the uniqueness of the result.

Once such criterion is that, when we compute the bulk two-point function for a given pair of operator insertions, we examine the identity block in both channels and choose the larger one. This can be justified in the following way. In a given channel each contribution to the two-point function from different blocks (operators) is positive
and if we take all corrections into account we should get the same result for both channels. 
Thus, the larger of the identity blocks in different channels is a better approximation.
In other words, in the channel with a smaller identity block contribution we would have a larger error if we neglect everything except for the identity block.

As a result,  the two-point function Equ.~(\ref{eq:2pt}) for a holographic BCFT on an UHP is given by
\begin{equation}
    \langle\bar{\Phi}_{n}(z_{L},\bar{z}_{L})\Phi_{n}(z_{R},\bar{z}_{R})\rangle_{\text{UHP}}=\max\left(\frac{\epsilon^{2d_{n}}}{|z_{L}-z_{R}|^{2d_{n}}},\frac{\mathcal{B}_{\Phi_{n}1}^{b}\mathcal{B}_{\Phi_{n}1}^{b}}{|z_{L}-z_{L}^{*}|^{d_{n}}|z_{R}-z_{R}^{*}|^{d_{n}}}\right) \, ,\label{eq:EEmax}
\end{equation}
where the first (second) term is the contribution from bulk (boundary) channel.
From this expression, we can immediately deduce the entanglement entropy for the TFD state of a holographic BCFT in flat background. However, we defer details to later sections where we consider the more general case of a curved background.
Nevertheless, we notice that the maximization prescription for the two-point function is equivalent to a \textit{minimization} prescription for the entanglement entropy (due to the minus sign in Eq.~\eqref{eq:defintion}), and this is consistent with the Ryu-Takayanagi (RT) prescription in holographic computations \cite{Ryu:2006bv,Ryu:2006ef}. We will in fact see that the result exactly matches the RT calculation.

\section{The Gravitational Calculation}\label{eq:gravcal}
\begin{figure}
    \centering
    \includegraphics[width=15cm]{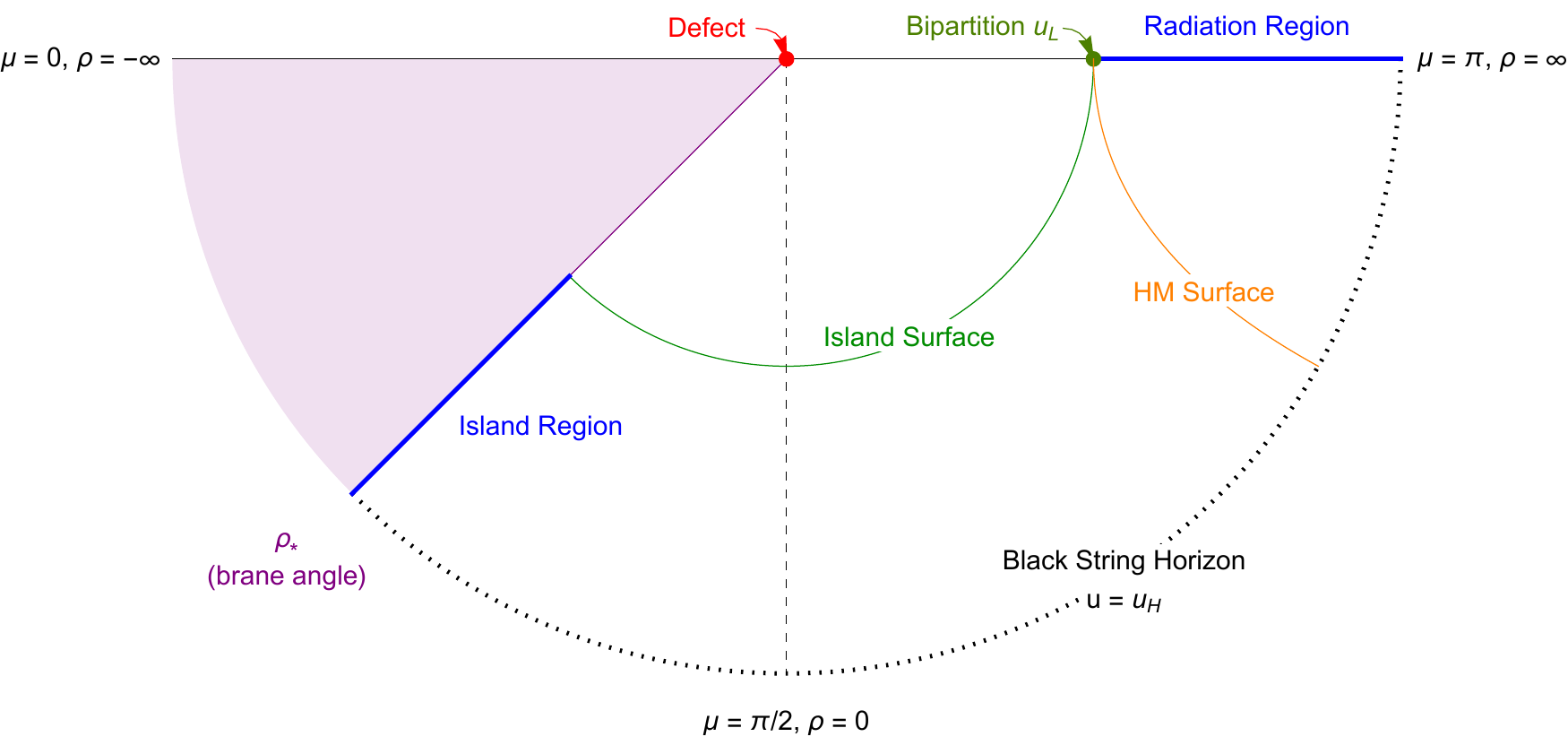}
    \caption{A cartoon representation of the ``left-side" thermofield double is shown here in the Poincar\'e half-plane. The light purple shaded region is excised from the AdS bulk by the KR brane at constant angular coordinate $\rho_*$. The convenience of the $\rho$ coordinate over the $\mu$ coordinate is clear here---we have conformally compactified the braneworld's infinite extra dimension into this finite diagram, so it now runs along the angular coordinate in the Poincar\'e half-plane and we identify $\mu=0, \frac{\pi}{2}, \pi$ with $\rho=-\infty, 0, \infty$, respectively. Also shown are the island and radiation regions, drawn in blue, and the island surface and HM surface, drawn in green and orange, respectively. The black string horizon forms a dotted arc at the coordinate $u=u_H$. For the eternal black string, we must also consider the ``right-side" thermofield double, which looks identical to the diagram above, except that we allow a bipartition for the radiation region at a different coordinate $u=u_R$, where we may have $u_L\neq u_R$.}
    \label{fig:LUP}
\end{figure}

The bulk geometry we are considering is the AdS$_3$ black string whose bulk geometry is given by $d=2$ of \eqref{eq:metricd}, with the metric
\begin{equation}
    ds^2=\frac{1}{u^2\sin^{2}\mu}\left[-\left(1-\frac{u}{u_{H}}\right)dt^2+\frac{du^2}{1-\frac{u}{u_{H}}}+u^2d\mu^2\right]\,.\label{eq:metricd2}
\end{equation}
For the sake of convenience we will use the following reparameterization
\begin{equation}
    \frac{1}{\sin\mu}=\cosh\rho,
\end{equation}
and the metric becomes
\begin{equation}
    ds^2=\cosh^{2}\rho \left[-\frac{(1-\frac{u}{u_{H}})}{u^2}dt^2+\frac{du^2}{u^2(1-\frac{u}{u_{H}})}\right]+d\rho^2\,.\label{eq:metric2}
\end{equation}
A complete picture of this geometric setup, illustrating the use of the $\rho$ coordinate and its correspondence with the $\mu$ coordinate, is presented in Figure \ref{fig:LUP}.

This geometry can be obtained using the embedding space formalism where the geometry is embedded as a codimension-one sub-manifold of a four dimensional Minkowski space
\begin{equation}
    ds^2=-dX_{0}^2-dX_{1}^2+dX_{2}^2+dX_{3}^2\,,
\end{equation}
with the following embedding equation
\begin{equation}
    X_{0}^1+X_{1}^2-X_{2}^2-X_{3}^2=1.
\end{equation}
The metric \eqref{eq:metric2} can be recovered using the following parameterization of the embedding equation
\begin{equation}
\begin{split}
        X_{0}&=\frac{2u_{H}-u}{u}\cosh\rho\,,\\
        X_{1}&=2\frac{\sqrt{u_{H}^2-uu_{H}}}{u}\sinh\frac{2\pi t}{\beta}\cosh\rho\,,\\
        X_{2}&=2\frac{\sqrt{u_{H}^2-uu_{H}}}{u}\cosh\frac{2\pi t}{\beta}\cosh\rho\,,\\
        X_{3}&=\sinh\rho\,,\label{eq:embedding1}
    \end{split}
\end{equation}
where $\beta=4\pi u_{H}$ is the inverse Hawking temperature. The advantage of using this embedding space formalism is that it is  easy to calculate the area of the Hartman-Maldacena surface without actually solving the minimal area (geodesic) differential equation. In this embedding, we can calculate the length $\ell$ of a geodesic between the coordinates $(X_0,X_1,X_2,X_3)$ and $(X_0',X_1',X_2',X_3')$ as
\begin{equation}
    \ell=\cosh^{-1}(X_0X_0'+X_1X_1'-X_2X_2'-X_3X_3').\label{eq:geolength}
\end{equation}

\begin{figure}
    \centering
    \includegraphics[width=10cm]{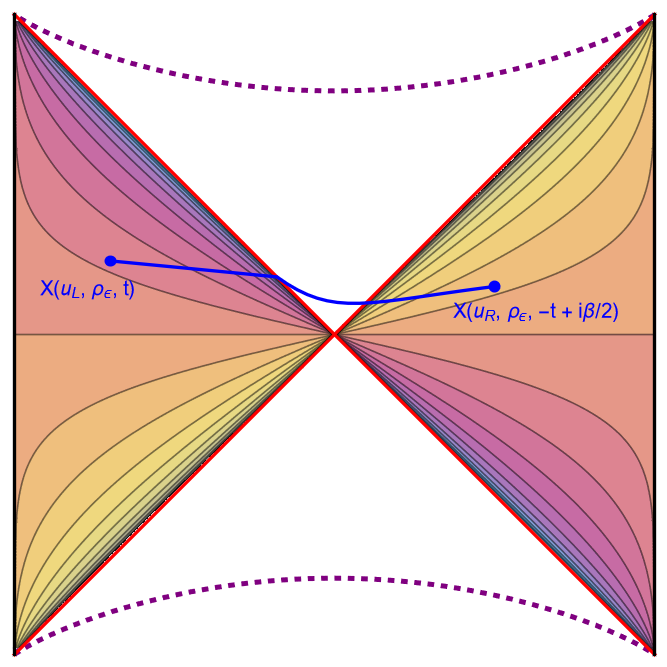}
    \caption{An AdS Penrose diagram shows a cartoon projection of the Hartman-Maldacena surface in blue, connecting our two bipartition points, on the left- and right-universe patches (left- and right-exterior). The Penrose diagram is shown as a projection onto a constant $\rho=\rho_\epsilon$ slice, where the purple dashed line is the singularity at $u=\infty$, the red diagonal lines are the event horizon at $u=u_H$, and the black borders on the left and right represent the defect at $u=0$. The color-gradient contours show the constant time-slices and reversed direction of the time-like Killing vector field on each side of the black hole. Notice that in this diagram, $u_L\neq u_R$, and the HM surface is not symmetric across the event horizon. Also note that the HM surface is not confined to this diagram since it generically varies in the coordinate $\rho$.}
    \label{fig:Penrose1}
\end{figure}

In contrast to the island surface, which starts at the bipartition and ends on the KR brane, the HM surface passes through an Einstein-Rosen bridge and ends at the right bipartition on the right-side thermofield double. In our $u-\rho$ coordinates, the left and right bipartitions lie at $(u,\rho)=(u_L,\infty)$ and $(u,\rho)=(u_R,\infty)$. To take advantage of the embedding \eqref{eq:embedding1}, we  introduce a regularization parameter $\rho_\epsilon$, which we will take to $\infty$ so that our bipartitions lie on the asymptotic boundary. Then, we may embed the left bipartition as
\begin{equation}
\begin{split}
        X_{0}^{L}&=\frac{2u_{H}-u_{L}}{u_{L}}\cosh\rho_{\epsilon}\,,\\
        X_{1}^{L}&=2\frac{\sqrt{u_{H}^2-u_{L}u_{H}}}{u_{L}}\sinh\frac{2\pi t}{\beta}\cosh\rho_{\epsilon}\,,\\
        X_{2}^{L}&=2\frac{\sqrt{u_{H}^2-u_{L}u_{H}}}{u_{L}}\cosh\frac{2\pi t}{\beta}\cosh\rho_{\epsilon}\,,\\
        X_{3}^{L}&=\sinh\rho_{\epsilon}\,,
    \end{split}
\end{equation}
and the right bipartition as
\begin{equation}
\begin{split}
        X_{0}^{R}&=\frac{2u_{H}-u_{R}}{u_{R}}\cosh\rho_{\epsilon}\,,\\
        X_{1}^{R}&=2\frac{\sqrt{u_{H}^2-u_{R}u_{H}}}{u_{R}}\sinh\frac{2\pi (-t+\frac{i\beta}{2})}{\beta}\cosh\rho_{\epsilon}\,,\\
        &=2\frac{\sqrt{u_{H}^2-u_{R}u_{H}}}{u_{R}}\sinh\frac{2\pi t}{\beta}\cosh\rho_{\epsilon}\,,\\
        X_{2}^{R}&=2\frac{\sqrt{u_{H}^2-u_{R}u_{H}}}{u_{R}}\cosh\frac{2\pi (-t+\frac{i\beta}{2})}{\beta}\cosh\rho_{\epsilon}\,,\\
        &=-2\frac{\sqrt{u_{H}^2-u_{R}u_{H}}}{u_{R}}\cosh\frac{2\pi t}{\beta}\cosh\rho_{\epsilon}\,,\\
        X_{3}^{R}&=\sinh\rho_{\epsilon}\,.
    \end{split}
\end{equation}
Note that for the right bipartition, we have taken the time coordinate $t\mapsto -t+\frac{i\beta}{2}$, corresponding to the reversal of the time-like Killing vector field on the other side of the black string horizon. For clarity, the situation is shown in Figure \ref{fig:Penrose1}.
Henceforth, we will use the substitutions
\begin{equation}
    \Delta_L=u_H-u_L,\quad \Delta_R=u_H-u_R
\end{equation}
where appropriate, to simplify and elucidate the physics in the following results. The area of the Hartman-Maldacena surface can now be readily calculated using \eqref{eq:geolength} as
\begin{equation}
\begin{split}
    A_{\text{HM}}&=\cosh^{-1}(X_{0}^{L}X_{0}^{R}+X_{1}^{L}X_{1}^{R}-X_{2}^{L}X_{2}^{R}-X_{3}^{L}X_{3}^{R})\,\\&=\cosh^{-1}\Big[{\frac{(2u_{H}-u_{L})(2u_{H}-u_{R})+4u_H\sqrt{\Delta_L\Delta_R}\cosh(\frac{4\pi t}{\beta})}{u_{L}u_{R}}\cosh^{2}(\rho_{\epsilon})-\sinh^{2}(\rho_{\epsilon})}\Big]\,.
    \end{split}
\end{equation}
Note the hyperbolic trig identities
\begin{align}
    \cosh^2(x)=\frac{e^{2x}}{4}+\frac{e^{-2x}}{4}+\frac{1}{2},\\
    \sinh^2(x)=\frac{e^{2x}}{4}-\frac{e^{-2x}}{4}+\frac{1}{2}.
\end{align}
Since $\rho_{\epsilon}$ is large, we can substitute $\cosh^2(x)\mapsto\frac{e^{2x}}{4}$, $\sinh^2(x)\mapsto\frac{e^{2x}}{4}$, and $\cosh^{-1}(x)\mapsto \log(2x)$. Following these substitutions, we obtain 
\begin{equation}
\begin{split}
    A_{\text{HM}}&=\log\Big[\frac{(2u_{H}-u_{L})(2u_{H}-u_{R})+4u_H\sqrt{\Delta_L\Delta_R}\cosh(\frac{4\pi t}{\beta})}{2u_{L}u_{R}}e^{2\rho_{\epsilon}}-\frac{e^{2\rho_{\epsilon}}}{2}\Big]\\&=\log\Big[\frac{(2u_{H}-u_{L})(2u_{H}-u_{R})+4u_H\sqrt{\Delta_L\Delta_R}\cosh(\frac{4\pi t}{\beta})}{2u_{L}u_{R}}-\frac{1}{2}\Big]+2\rho_{\epsilon}\\
    &=\log\Big[\frac{u_H}{u_Lu_R}\left(\Delta_L+\Delta_R+2\sqrt{\Delta_L\Delta_R}\cosh(\frac{4\pi t}{\beta})\right)\Big]+2\rho_{\epsilon}\,,
    \end{split}\label{eq:HMsurfacearea}
\end{equation}
and using the Ryu-Takayanagi formula we get the associated entanglement entropy
\begin{equation}
\begin{split}
S_{\text{HM}}&=\frac{A_{\text{HM}}}{4G_{3}}\\&=\frac{c}{6}\log\Big[\frac{u_H}{u_Lu_R}\left(\Delta_L+\Delta_R+2\sqrt{\Delta_L\Delta_R}\cosh(\frac{4\pi t}{\beta})\right)\Big]+\frac{c}{3}\rho_{\epsilon}\,,\label{eq:gravityEEHM}
\end{split}
\end{equation}
where we used the Brown-Henneaux central charge $c=\frac{3}{2G_{3}}$. In the special case $u=u_L=u_R$, using $\Delta_u=u_H-u$, this reduces to
\begin{equation}
    S_\text{HM}=\frac{c}{6}\log\Big[\frac{2u_H\Delta_u}{u^2}\left(1+\cosh(\frac{4\pi t}{\beta})\right)\Big]+\frac{c}{3}\rho_{\epsilon}\,.\label{eq:SHMGravityu}
\end{equation}
We now compare this result to the pair of minimal island surfaces crossing from the bipartitions $u_L$ and $u_R$ to their respective physical branes at $\rho=\rho_*$. In general, we  compute the area of the island surface
\begin{equation}
    A_{\text{island}}=\underset{\text{min. island surface}}{\int} ds
\end{equation}
and rewrite $ds$ in terms of $d\rho$ using the metric in \eqref{eq:metric2}
\begin{equation}
    ds=d\rho\sqrt{1+\frac{\cosh^2\rho}{u(\rho)^2\left(1-\frac{u(\rho)}{u_H}\right)}u'(\rho)^2}.\label{eq:dsdrho}
\end{equation}
This is clearly minimized for $u'(\rho)=0\implies u\text{ constant}$, so the combined area for the pair of island surfaces reads
\begin{equation}
    \begin{split}
        A_{\text{island}}=2\int_{\rho_{*}}^{\rho_{\epsilon}} d\rho=2(\rho_{\epsilon}-\rho_{*})\,,
    \end{split}
\end{equation}
where $\rho_*$ is the location of the physical brane and the entanglement entropy calculated by the island surface is
\begin{equation}
\begin{split}
    S_{\text{island}}=\frac{A_{\text{island}}}{4G_{3}}=-\frac{c}{3}\rho_{*}+\frac{c}{3}\rho_{\epsilon}\,,\label{eq:gravEEisl}
    \end{split}
\end{equation}
%where we matched the brane location $\rho_{*}$ with the boundary entropy $\log g_{b}$ defined in Equ.~(\ref{eq:bdyentropy}) as $-2\rho_{*}=\frac{12}{c}\log g_{b}$. 
According to the RT proposal the actual entanglement entropy is the minimum of $S_{\text{HM}}$ and $S_{\text{island}}$
\begin{equation}
    S=\min(S_{\text{HM}},S_{\text{island}})\,.\label{eq:entropymin}
\end{equation}

\section{The Field Theory Calculation}
In this section, we apply the techniques developed in Sec.\ref{sec:reviewCFT} to provide a direct field theory computation of the entanglement entropy that we computed holographically in the previous section.
\subsection{Useful Geometries}
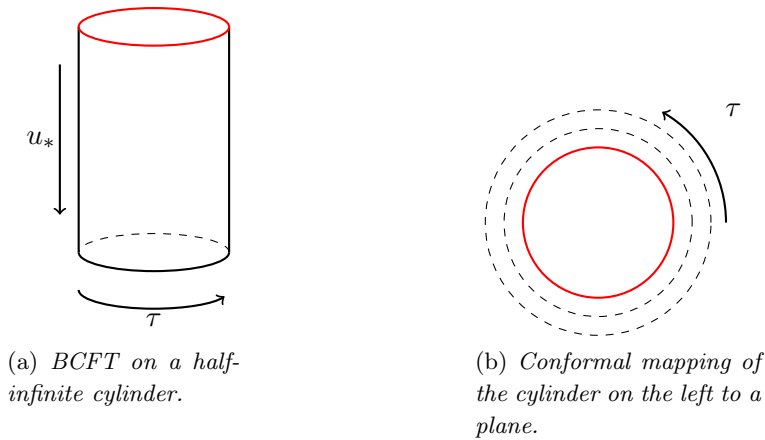
\begin{figure}
\centering
\subfloat[\textit{BCFT on a half-infinite cylinder.}\label{cylinder}]
{
\begin{tikzpicture}[scale=0.5]
\draw[-,thick,black] (0,-3) to (0,3);
\draw[-,thick,black] (4,-3) to (4,3);
\draw[-,thick,black](4,-3) arc (0:-180:2 and 0.5);
\draw[-,dashed,black](4,-3) arc (0:180:2 and 0.5);
\draw[-,thick,red] (4,3) arc (0:360:2 and 0.5);
\draw[->,thick,black] (-0.5,2) to (-0.5,-2);
\draw[->,thick,black](0,-4) arc (-180:-20:2 and 0.5);
\node at (-1,0) {\textcolor{black}{$u_{*}$}};
\node at (2,-4.8)
{\textcolor{black}{$\tau$}};
\end{tikzpicture}\label{pic:cylinder}
}
\hspace{3cm}
\subfloat[\textit{Conformal mapping of the cylinder on the left to a plane.} \label{plane}]
{
\begin{tikzpicture}[scale=0.5]
\draw[-,thick,red] (2,0) arc (0:360:2);
\draw[-,dashed,black] (2.5,0) arc (0:360:2.5);
\draw[-,dashed,black] (3,0) arc (0:360:3);
\draw[->,thick,black] (3.4,0) arc (0:60:3.4);
\node at (3.6,3)
{\textcolor{black}{$\tau$}};
\end{tikzpicture}\label{pic:plane}
}
\caption{\small{\textit{a) The geometry of eq.~(\ref{eq:cylinder}) is (conformally) a half-infinitely long cylinder. The red circle is the place of the conformal boundary $u_{*}=0$.
b) The cylinder is now conformally transformed to a plane by eq.~(\ref{eq:cyntopl}) where the image of the cylinder covers only the region outside of the red circle. The time coordinate $\tau$ is mapped to the angular coordinate on this plane, constant $u_{*}$ slices are concentric circles and the red circle is the conformal boundary. }}}
\label{demon}
\end{figure}

Since the field theory is living on a curved background, to compute the entanglement entropy on the field theory side, we first study the geometry on the field side. The geometry is given by the boundary geometry ($\mu=\pi$) of Eq.~(\ref{eq:metricd}). This geometry is described by the following metric

\begin{equation}
    ds^{2}=-\frac{1}{u^{2}}(1-\frac{u^{d-1}}{u_{H}^{d-1}})dt^2+\frac{du^2}{u^{2}(1-\frac{u^{d-1}}{u_{H}^{d-1}})}+\frac{d\vec{x}^{2}_{d-2}}{u^2}\,,
\end{equation}
which is an AdS$_{d}$ planar black hole for a generic $d$ and we have two copies of such geometry corresponding to the two asymptotic boundaries of Eq.~(\ref{eq:metricd}). We consider the special case $d=2$ where the metric is
\begin{equation}
    ds^{2}=-\frac{1}{u^2}(1-\frac{u}{u_{H}})dt^2+\frac{du^2}{u^{2}(1-\frac{u}{u_{H}})}\,,\label{eq:metricft}
\end{equation}
and the conformal boundary of the 2d BCFT is located at $u=0$. This metric can be put into the conformally flat form by the following coordinate transform
\begin{equation}
    ds^{2}=\Omega(u_{\star})^{2}\big[-dt^{2}+du_{\star}^{2}\big]\,,\quad u_{\star}=-u_{H}\log(1-\frac{u}{u_{H}})\,,\label{eq:metriccylinder}
\end{equation}
where $u_{\star}\in(0,\infty)$ and the conformal factor is 
\begin{equation}
    \Omega(u_{\star}(u))=\frac{1}{u}\sqrt{1-\frac{u}{u_{H}}}\,.
\end{equation}
To incorporate the fact that we have two asymptotic boundaries and a finite temperature in the path integral language, we will consider the Euclidean version of the geometry
\begin{equation}
    ds^{2}=\Omega(u_{\star})^{2}\big[d\tau^{2}+du_{\star}^2\big]\,,\quad \tau\sim \tau+\beta\,,\label{eq:cylinder}
\end{equation}
where the zero-time slices of the two asymptotic boundaries of eq.~(\ref{eq:metric2}) are the $\tau=0$ and $\tau=\frac{\beta}{2}$ slices respectively and $\tau=it$. Therefore we have a cylindrical geometry the perimeter of whose cross section is $\beta$ and it is half infinitely long (see Fig.~\ref{pic:cylinder}).

To study the conformal properties of this geometry we can use complex coordinates
\begin{equation}
    z=u_{\star}+i\tau\,,\quad\bar{z}=u_{\star}-i\tau\,.\label{eq:z}
\end{equation}
This geometry can be conformally mapped to the plane with a disk $w\bar{w}\leq u_{H}^2$ removed (i.e. the conformal boundary is now the circle $w\bar{w}=u_{H}^2$) by the following map
\begin{equation}
    w=u_{H}e^{\frac{z}{2u_{H}}}\,,\label{eq:cyntopl}
\end{equation}
and in this new coordinate the metric %transforms
is
\begin{equation}
    ds^{2}=\Omega(u_{\star})^{2}dzd\bar{z}=4\Omega(u_{\star})^{2}e^{-\frac{u_{\star}}{u_{H}}}dwd\bar{w}\,.\label{eq:metricPlane}
\end{equation}
This geometry can be understood as a path integral preparing a thermofield double state (TFD) of two half-space BCFTs \cite{Sully:2020pza,Geng:2021iyq} if we use $\arg(w)$ (i.e. $\tau$) as the time-coordinate and it is in the vacuum state of two half-space BCFTs if we instead use $\Im(w)$ as the time-coordinate. 

Similar to Equ.~(\ref{eq:TFDmap1}), we can further conformally map the geometry to the upper-half-plane (UHP) by
\begin{equation}
    w=\frac{u_{H}}{v-\frac{i}{2}}-i u_{H}\,,\label{eq:pltoUHP}
\end{equation}
where the conformal boundary is mapped to the real axis $v-\bar{v}=0$ and the metric transforms to
\begin{equation}
    ds^{2}=4\Omega(u_{\star})^{2}e^{-\frac{u_{\star}}{u_{H}}}u_{H}^{2}(e^{\frac{z}{2u_{H}}}+i)^{2}(e^{\frac{\bar{z}}{2u_{H}}}-i)^{2}dvd\bar{v}\,.\label{eq:metricUHP}
\end{equation}

\begin{figure}
    \centering
    \begin{tikzpicture}
    \draw[fill=gray, draw=none, fill opacity = 0.1] (-3,0) to (3,0) to (3,2) to (-3,2);
    \draw[-,thick,red] (-3,0) to (3,0);
    \end{tikzpicture}
    \caption{The plane with a conformal boundary Fig.~(\ref{pic:plane}) can be further conformally mapped to an upper-half-plane (UHP) by Eq.~(\ref{eq:pltoUHP}). The conformal boundary is mapped now to the real axis of the plane.}
    \label{fig:UHP}
\end{figure}
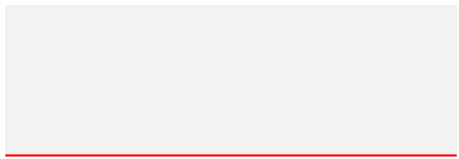

\subsection{Computation of the Entanglement Entropy}
To compute the entanglement entropy, we first consider the simplest bipartition: $u_{L}=u_{R}=u$. As we reviewed in Sec.~\ref{sec:reviewCFT}, the entanglement entropy can be computed by first calculating the two-point function $\langle\Phi_{n}(u_{R},t)\Phi_{n}(u_{L},t)\rangle$ of the twist operator $\Phi_{n}(u,t)$ in the geometry of Eq.~(\ref{eq:metriccylinder}). Since we are studying a holographic BCFT \cite{Rozali:2019day,Sully:2020pza}, we will use the vacuum dominance to compute this two-point function. The result will be similar to Equ.~(\ref{eq:EEmax}). We will see that the boundary channel will reproduce the result of the island surface and the bulk channel will reproduce the result of the HM surface in the gravity side calculation in Sec.~\ref{eq:gravcal}.

\subsubsection{Boundary Channel}

Let's first consider the boundary channel. As we reviewed in Sec.~\ref{sec:reviewCFT}, this is in a sense a disconnected channel and equivalently we are just computing the product of two one-point functions $\langle\phi_{n}(u_{L},t)\rangle_{b}\langle\phi_{n}(u_{R},t)\rangle_{b}$ for operators $\phi_{n}(u_{L(R)},t)$ near the boundary. We know such one-point functions when the field theory is on an flat UHP. However, in the current case we can map the BCFT to a conformally flat UHP described Eq.~(\ref{eq:metricUHP}) and as opposed to the flat case we have to further dress the correlator by the conformal factor. The result is given by
\begin{equation}
\begin{split}
    \langle\Phi_{n}(u,t)\rangle_{b}&=2^{-\Delta_{n}}e^{\frac{u_{\star}}{2u_{H}}\Delta_{n}}u_{H}^{-\Delta_{n}}(e^{\frac{z}{2u_{H}}}+i)^{-\Delta_{n}}(e^{\frac{\bar{z}}{2u_{H}}}-i)^{-\Delta_{n}}\Omega(u_{\star}(u))^{-\Delta_{n}}\langle\Phi_{n}(v,\bar{v})\rangle^{b}_{\text{Flat UHP}}\\&=\frac{2^{-\Delta_{n}}e^{\frac{u_{\star}}{2u_{H}}\Delta_{n}}u_{H}^{-\Delta_{n}}(e^{\frac{z}{2u_{H}}}+i)^{-\Delta_{n}}(e^{\frac{\bar{z}}{2u_{H}}}-i)^{-\Delta_{n}}\Omega(u_{\star}(u))^{-\Delta_{n}}\mathcal{B}_{\Phi_{n}1}^{b}}{(v-\bar{v})^{\Delta_{n}}}\\&=\frac{\mathcal{B}_{\Phi_{n}1}^{b}}{2^{\Delta_{n}}}\,,
    \end{split}
\end{equation}
where we follow the notation we introduced in Sec.~\ref{sec:reviewCFT} that $\mathcal{B}_{\Phi_{n}1}^{b}$ denotes the coefficient of the identity operator $\mathbf{1}$ in the boundary operator expansion (BOE) of $\Phi_{n}$. As a result, the two point function in the boundary channel is given by
\begin{equation}
    \langle\Phi_{n}(u_{L},t_{L})\Phi_{n}(u_{R},t_{R})\rangle_{\text{bdy}}=\frac{\mathcal{B}_{\Phi_{n}1}^{b}\mathcal{B}_{\Phi_{n}1}^{b}}{4^{\Delta_{n}}}\,.
\end{equation}
Therefore the entanglement entropy is
\begin{equation}
\begin{split}
    S_{\text{bdy}}&=\lim_{n\rightarrow1}\frac{1}{1-n}\log\langle\Phi_{n}(u_{L},t_{L})\Phi_{n}(u_{R},t_{R})\rangle_{\text{bdy}}\\&=2\log g_{b}+\frac{c}{3}\log(\frac{2}{\epsilon})\,,\label{eq:fieldEEbdy}
    \end{split}
\end{equation}
where we have used Equ.~(\ref{eq:bdyentropy}) to the BOE coefficient by the boundary entropy term. 

\subsubsection{Bulk Channel}
Now let's consider the bulk channel. Applying vacuum dominance, we are just computing the two-point function of the primary operator as if there is no boundary. We know the result for such a two-point function if the field theory is on a flat plane. In our case we can conformally map the geometry into a conformally flat plane as in Equ.~{\ref{eq:metricPlane}}. The result of the two-point function is the same as that in the flat plane Equ.~(\ref{eq:bulkchannel}) case, however with the conformal factor $\Omega(u_{*})$ properly dressed. It is given by
\begin{equation}
    \begin{split}
        \langle\Phi_{n}(u_{L},t_{L})\Phi_{n}(u_{R},t_{R})\rangle_{\text{bulk}}&=\langle\Phi_{n}(u,-t+i\frac{\beta}{2})\Phi_{n}(u,t)\rangle_{\text{bulk}}\\&=4^{-\Delta_{n}}e^{\frac{u_{\star}\Delta_{n}}{2u_{H}}}\Omega(u)^{-\Delta_{n}}e^{\frac{u_{\star}\Delta_{n}}{2u_{H}}}\Omega(u)^{-\Delta_{n}}\langle\Phi_{n}(u,-t+i\frac{\beta}{2})\Phi_{n}(u,t)\rangle_{\text{Flat Plane}}\\&=\frac{4^{-\Delta_{n}}\Omega(u)^{-2\Delta_{n}}e^{\frac{u_{\star}\Delta_{n}}{u_{H}}}\epsilon^{2\Delta_{n}}}{\abs{w_{L}-w_{R}}^{2\Delta_{n}}}\,, \label{eq:2ptbulk}
    \end{split}
\end{equation}
where we followed Equ.~(\ref{eq:bulkchannel}) to put a cutoff $\epsilon$ for later convenience. Then we have the entanglement entropy calculated by the bulk channel
\begin{equation}
\begin{split}
    S_{\text{bulk}}&=-\frac{1}{n-1}\lim_{n\rightarrow1}\tr\rho^n\\&=\lim_{n\rightarrow1}\frac{1}{1-n}\log\langle\Phi_{n}(u_{L},t_{L})\Phi_{n}(u_{R},t_{R})\rangle_{\text{bulk}}\\&=\frac{c}{6}\log\Bigg[\frac{4u_{H}^2}{\epsilon^{2}u^{2}}\abs{e^{\frac{i\tau}{2u_{H}}}\sqrt{1-\frac{u}{u_{H}}}+e^{-\frac{i\tau}{2u_{H}}}\sqrt{1-\frac{u}{u_{H}}}}^2\Bigg]\\
    &=\frac{c}{6}\log\left[\frac{2u_H(u_H-u)}{u^2}\left(1+\cosh\frac{4\pi t}{\beta}\right)\right]+\frac{c}{3}\log(\frac{2}{\epsilon})\,,
    %\\
    %&=\frac{c}{3}\log\Bigg[\frac{2u_{H}}{u\epsilon}\sqrt{1-\frac{u}{u_{H}}}\cosh\frac{2\pi t}{\beta}\Bigg]
    \end{split}
\end{equation}
where we used Equ.~(\ref{eq:cyntopl}), Equ.~(\ref{eq:z}) and Equ.~(\ref{eq:metriccylinder}) to express $\omega_{L}$ and $\omega_{R}$ in Equ.~(\ref{eq:2ptbulk}) in terms of $u_{L}=u_{R}=u$ and $\tau$ and we transformed back to the Lorenzian signature $t=-i\tau$ at the end. We can rewrite this in the $\Delta_u=u_H-u$ convention as
\begin{equation}
    S_\text{bulk}=\frac{c}{6}\log\left[\frac{2u_H\Delta_u}{u^2}\left(1+\cosh\frac{4\pi t}{\beta}\right)\right]+\frac{c}{3}\log(\frac{2}{\epsilon})\,.\label{eq:fieldEEbulk}
\end{equation}
% \begin{equation}
%     S_{\text{bulk}}=\frac{c}{3}\log\left[\frac{2u_H}{y\epsilon}\right]
% \end{equation}
Hence comparing the field theory results Eq.~(\ref{eq:fieldEEbdy}) and Eq.~(\ref{eq:fieldEEbulk}) with the gravity side calculations Eq.~(\ref{eq:gravEEisl}) and Eq.~(\ref{eq:gravityEEHM}) (with $u_{L}=u_{R}=u$), we can see that the field theory results precisely match the gravity side calculations if we use the following identifications
\begin{equation}
\log g_{b}=-\frac{c}{6}\rho_{*}\,,\quad\epsilon=2e^{-\rho_{\epsilon}}\,.\label{eq:cutoffid}
\end{equation}

Now for the most general case that $u_{L}\neq u_{R}$ we have 
\begin{equation}
    \begin{split}
        \langle\Phi_{n}(u_{L},t_{L})\Phi_{n}(u_{R},t_{R})\rangle_{\text{bulk}}&=\langle\Phi_{n}(u,-t+i\frac{\beta}{2})\Phi_{n}(u,t)\rangle_{\text{bulk}}\\&=4^{-\Delta_{n}}e^{\frac{u_{\star L}\Delta_{n}}{2u_{H}}}\Omega(u_{L})^{-\Delta_{n}}e^{\frac{u_{\star R}\Delta_{n}}{2u_{H}}}\Omega(u_R)^{-\Delta_{n}}\\
        &\qquad\qquad\times\langle\Phi_{n}(u_{L},-t+i\frac{\beta}{2})\Phi_{n}(u_{R},t)\rangle_{\text{Plane}}\\&=\frac{4^{-\Delta_{n}}e^{\frac{u_{\star L}\Delta_{n}}{2u_{H}}}\Omega(u_{L})^{-\Delta_{n}}e^{\frac{u_{\star R}\Delta_{n}}{2u_{H}}}\Omega(u_R)^{-\Delta_{n}}\epsilon^{2\Delta_{n}}}{\abs{w_{L}-w_{R}}^{2\Delta_{n}}}\,.
    \end{split}
\end{equation}
Therefore, the entanglement entropy calculated by the bulk channel is
\begin{equation}
\begin{split}
    S_{\text{bulk}}&=\lim_{n\rightarrow1}\frac{1}{1-n}\log\langle\Phi_{n}(u_{L},t_{L})\Phi_{n}(u_{R},t_{R})\rangle_{\text{bulk}}\\&=\frac{c}{6}\log\Bigg[\frac{4u_{H}^2}{\epsilon^{2}u_{L}u_{R}}\abs{e^{-\frac{i\tau}{2u_{H}}}\sqrt{1-\frac{u_{R}}{u_{H}}}+e^{\frac{i\tau}{2u_{H}}}\sqrt{1-\frac{u_{L}}{u_{H}}}}^2\Bigg]\\&=\frac{c}{6}\log\Bigg[\frac{4u_{H}^2}{\epsilon^{2}u_{L}u_{R}}\big((1-\frac{u_{R}}{u_{H}})+(1-\frac{u_{L}}{u_{H}})+2\sqrt{1-\frac{u_R}{u_H}}\sqrt{1-\frac{u_L}{u_H}}\frac{e^{\frac{i\tau}{u_{H}}}+e^{-\frac{i\tau}{u_{H}}}}{2}\big)\Bigg]\,\\
    % &=\frac{c}{6}\log\Big[\frac{(2u_{H}-u_{L})(2u_{H}-u_{R})}{2u_{L}u_{R}}+\frac{4\sqrt{u_{H}^{2}-u_{H}u_{L}}\sqrt{u_{H}^2-u_{H}u_{R}}}{2u_{L}u_{R}}\cosh(\frac{4\pi t}{\beta})-\frac{1}{2}\Big]-\frac{c}{3}\log\epsilon\\
    &=\frac{c}{6}\log\Big[\frac{4u_H^2-2u_Hu_L-2u_Hu_R}{2u_{L}u_{R}}+\frac{4u_H\sqrt{\Delta_L\Delta_R}}{2u_{L}u_{R}}\cosh(\frac{4\pi t}{\beta})\Big]+\frac{c}{3}\log(\frac{2}{\epsilon})
    \\
    &=\frac{c}{6}\log\Big[\frac{u_H}{u_Lu_R}\left(\Delta_L+\Delta_R+2\sqrt{\Delta_L\Delta_R}\cosh(\frac{4\pi t}{\beta})\right)\Big]+\frac{c}{3}\log(\frac{2}{\epsilon})\,,\label{eq:fieldEEbulkgen}
    \end{split}
\end{equation}
which again precisely matches the gravity side calculation Equ.~(\ref{eq:gravityEEHM}) under the identification of the cutoffs Equ.~(\ref{eq:cutoffid}).

\section{Limits and Page Curve Behavior}
%I think I fixed this one
Having shown that the gravitational and the field theory calculations of $S_\text{island}=S_\text{bdy}$ and $S_\text{HM}=S_\text{bulk}$ match, we can now describe the entanglement entropy given by the minimum in \eqref{eq:entropymin}. With these analytical results, we will perform an analysis similar to \cite{Geng:2020fxl} on the nature of the Page time and Page angle in our $\text{AdS}_3$ setup.

First, we note that in the limit $t\to 0$, the area of the HM surface, Equ. (\ref{eq:HMsurfacearea}), goes to
\begin{equation}
    A_{\text{HM}}=\log\Big[\frac{u_H}{u_Lu_R}\left(\sqrt{\Delta_L}+\sqrt{\Delta_R}\right)^2\Big]+2\rho_{\epsilon}\,.
\end{equation}
The Page time $t_{\text{Page}}$ can be calculated by
\begin{equation}
    S_{\text{HM}}(t_{\text{Page}})=S_{\text{island}}\,,
\end{equation}
and the result is
\begin{equation}
    t_{\text{Page}}=\frac{\beta}{4\pi}\cosh^{-1}\Bigg[\frac{e^{-2\rho_*}u_Lu_R-u_H(\Delta_L+\Delta_R)}{2u_H\sqrt{\Delta_L\Delta_R}}\Bigg]\,.
\end{equation}
For the special case $u_{L}=u_{R}=u$, and taking the convention $\Delta_u=u_H-u$, we have
\begin{equation}
    t_{\text{Page}}=\frac{\beta}{4\pi}\cosh^{-1}\left(\frac{e^{-2\rho_*}u^2}{2u_H\Delta_u}-1\right)\,.
\end{equation}
The Page angle $\rho_\text{Page}$ describes the brane angle at which a non-trivial Page curve arises ($t_\text{Page}>0$) and can be calculated by
\begin{equation}
    S_{\text{HM}}(t=0)=S_{\text{island}}(\rho_*)\,.
\end{equation}
The result is
\begin{equation}
    \rho_\text{Page}=\frac{1}{2}\log\left[\frac{u_Lu_R}{u_H\left(\sqrt{\Delta_L}+\sqrt{\Delta_R}\right)^2}\right]\,.\label{eq:PageAngle}
\end{equation}
with the special case $u=u_L=u_R$ given by
\begin{equation}
    \rho_\text{Page}=\frac{1}{2}\log\left[\frac{u^2}{4u_H\Delta_u}\right]\,.
\end{equation}
Here, the special case $u=u_L=u_R$ behaves similarly to the $u_L\neq u_R$ case---the Page angle increases monotonically in either variable. Then, let us analyze the special case. In particular, we plot the dependence of the Page angle on $u$ in Fig. \ref{fig:page angle u plot}. We can calculate the point at which the Page angle is 0, corresponding to a tensionless brane. The result is
\begin{equation}
    u_{\rho=0}=(2\sqrt{2}-2)u_H\,.
\end{equation}

This plot  also acts as a phase diagram of sorts---for a given bipartition location $u$, placing the KR brane at an value of $\rho$ below this curve leads to a non-trivial Page time, since the HM surface initially dominates. Above this curve, the island surface is always minimal, and dominates for all time. Note that for $u>u_{\rho=0}$, any positive-tension brane ($\rho_*<0$) will yield a non-trivial Page curve.
\begin{figure}[hb]
    \centering
    \includegraphics[width=11cm]{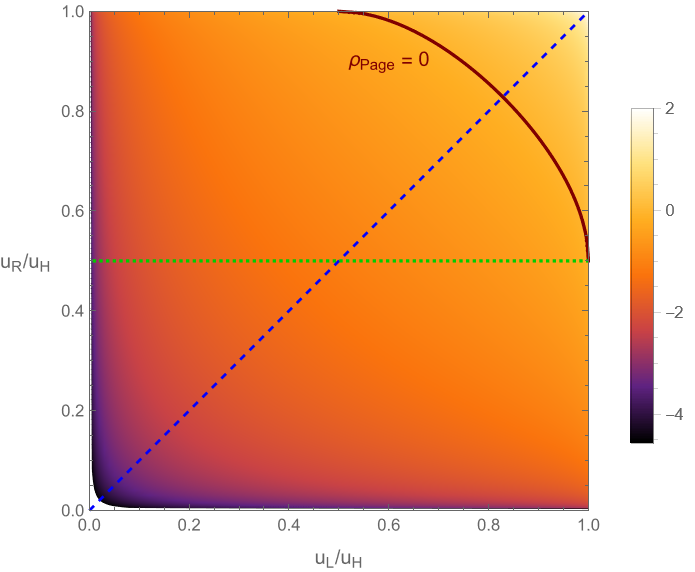}
    \caption{A density plot of the Page angle $\rho_{\text{Page}}$ with respect to both $u_L$ and $u_R$. The contour with $\rho_{\text{Page}}=0$ is shown in maroon as a point of reference. Drawn in dashed blue and dotted green are the slices of the parameter space shown in figures \ref{fig:page angle u plot} and \ref{fig:PageAngleuL}, respectively.}
    \label{fig:PageAngleuLuR}
\end{figure}

\begin{figure}[htp]
    \centering
    \includegraphics[width=12cm]{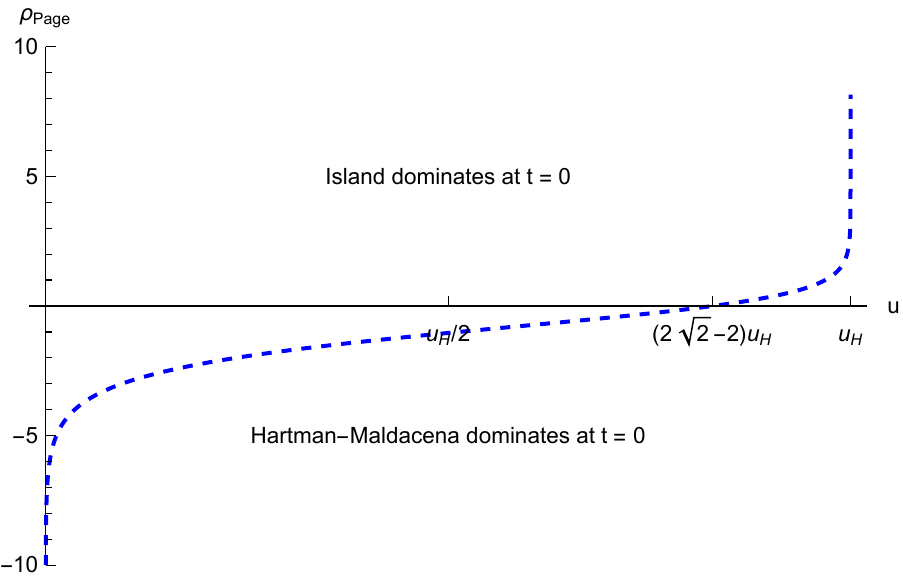}
    \caption{A plot of the Page angle $\rho_\text{Page}$ in the special case $u=u_L=u_R$. The point at which the Page angle gives the tensionless brane $\rho=0$ is marked on the horizontal axis, at $u=(2\sqrt{2}-2)u_H$. The regions in which the island and Hartman-Maldacena surfaces initially dominate the entropy calculation are marked. This plot is equivalent to plotting the density along a diagonal slice of Figure \ref{fig:PageAngleuLuR} where $u_L=u_R$.}
    \label{fig:page angle u plot}
\end{figure}

\begin{figure}[h]
    \centering
    \includegraphics[width=12cm]{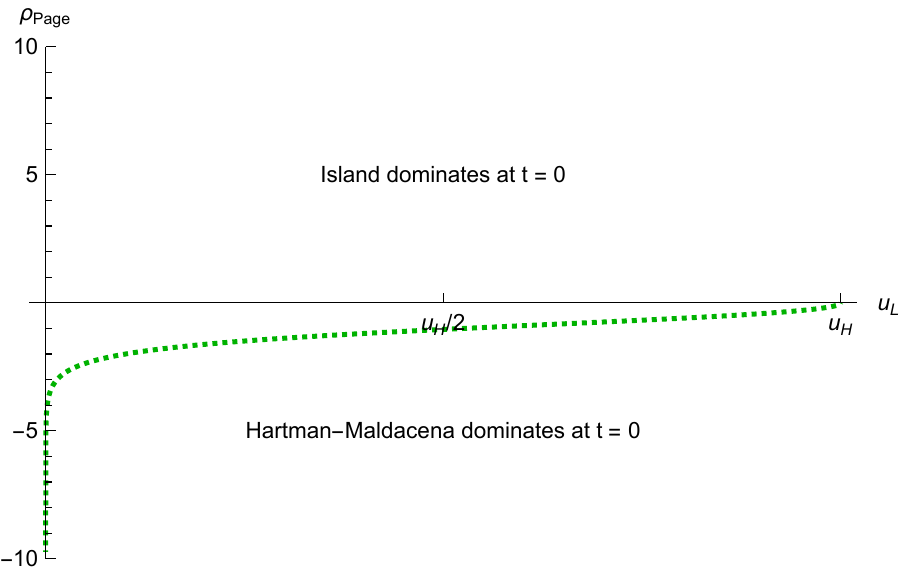}
    \caption{A plot of the Page angle $\rho_\text{Page}$ in the special case $u_R=\frac{u_H}{2}$. Note that the horizontal axis here now refers to the location of $u_L$. This plot is equivalent to plotting the density along a horizontal slice of Figure \ref{fig:PageAngleuLuR} at $u_R=\frac{u_H}{2}$.}
    \label{fig:PageAngleuL}
\end{figure}

We can also plot the Page angle as a function of $u_L$ and $u_R$ using a density plot, as shown in Figure \ref{fig:PageAngleuLuR}. This yields a somewhat surprising result---even for very different bipartition locations $u_L$ and $u_R$, the Page angle is roughly the same (notice the large orange region in the figure). It changes dramatically only when either of $u_R$ and $u_L$ are very close to $0$ or both are close to $u_H$. Figure \ref{fig:PageAngleuL} illustrates how the phase diagram changes when we fix one bipartition location (this time specifically at $\frac{u_H}{2}$)---the Page angle is very stable for a wide range of values $u_L$, blowing up only when the bipartition is very close to the defect. Following this analysis, we see that taking just one bipartition to $u_H$ gives us
\begin{equation}
    \lim_{u_R\to u_H}\rho_{\text{Page}}=\frac{1}{2}\log\left[\frac{u_L}{\Delta_L}\right]=\frac{1}{2}\log\left[\frac{u_L}{u_H-u_L}\right].
\end{equation}
In this limit, we also see that $u_L=\frac{u_H}{2}$ yields a Page angle of zero. Furthermore, we see from the contour $\rho_{\text{Page}}=0$ that if either $u_L$ or $u_R$ is less than $\frac{u_H}{2}$, there always exists a a configuration of $u_L$ and $u_R$ such that $\rho_\text{Page}<0$. Conversely, Page angles greater than 0 (a brane at such an angle would have negative tension) only arise when at least one of $u_L$ or $u_R$ is greater than $\frac{u_H}{2}$.

\begin{figure}
    \centering
    \includegraphics[width=11cm]{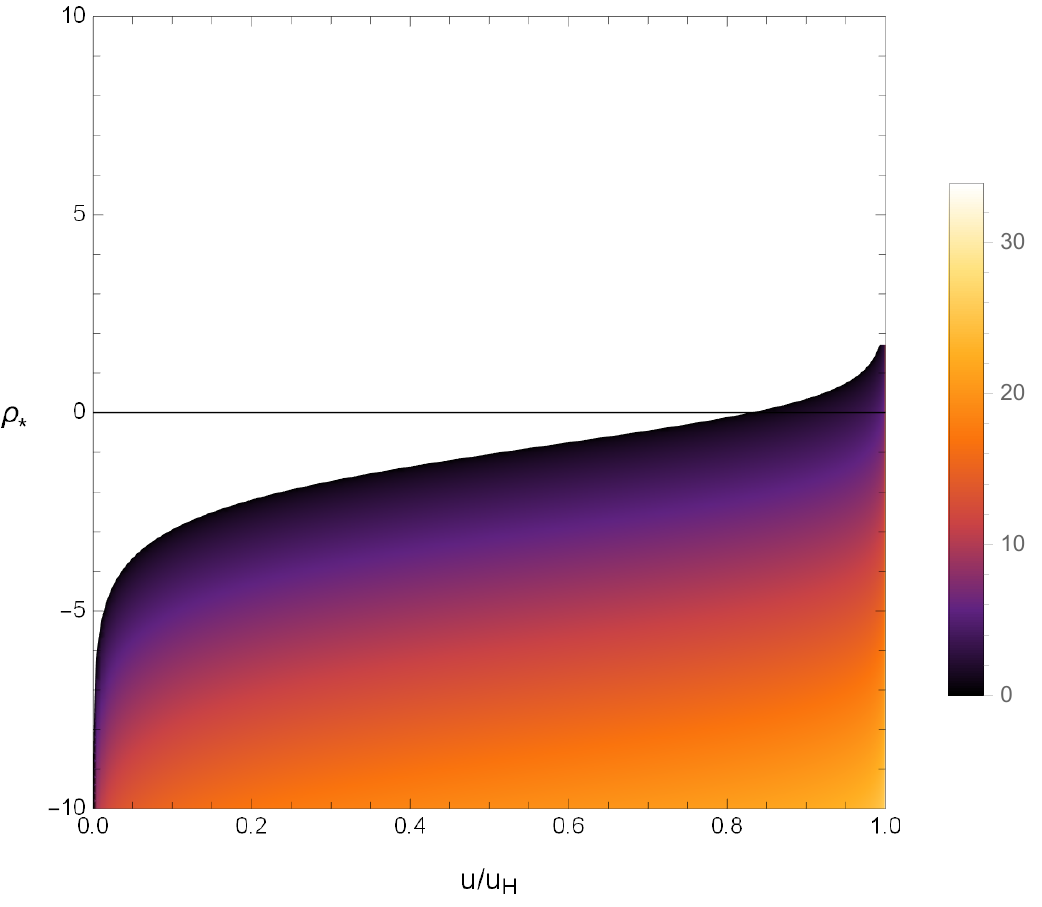}
    \caption{A density plot of the Page time $t_{\text{Page}}$, in the special case $u=u_L=u_R$. The plot cuts off at $t_{\text{Page}}=0$, at the boundary given by the Page angle plot in Fig. \ref{fig:page angle u plot}.}
    \label{fig:PageTimeDensity}
\end{figure}

To see the Page time more explicitly, we examine the density plot in Figure \ref{fig:PageTimeDensity}. Here we see that constant-time contours trace out plots similar to \ref{fig:page angle u plot} (which is, of course, the limiting edge of the density plot at $t_{\text{Page}}=0$). From this, we conclude that the Page time is roughly the same across a large range of $u$ coordinates, changing dramatically only when $u$ is close to $0$ or $u_H$. This is consistent with and reinforces our findings that the Page curve behavior of our setup is not strongly influenced by small or medium changes in the bipartition location, except when it is close to either the event horizon or the defect.

\section{Conclusion}

In this paper, we constructed a lower dimensional analytical model for a holographic BCFT living on a black hole background. The BCFT lives on a two-dimensional eternal AdS Schwarzschild geometry with conformal boundary conditions imposed on its asymptotic boundaries. The dual bulk geometry is an AdS$_3$ black string with an embedded Karch-Randall brane \cite{Karch:2000ct,Karch:2000gx}. This is a setup that satisfies the AdS/BCFT correspondence \cite{Fujita:2011fp}. 

We studied entanglement entropy for a two-sided bipartition for the BCFT. This bipartition is known to capture certain dynamical aspects of the system \cite{Hartman:2013qma} and is relevant to the recent Page curve calculation in the Karch-Randall braneworld \cite{Ling:2020laa,KumarBasak:2020ams, Emparan:2020znc,Caceres:2020jcn,Caceres:2021fuw,Deng:2020ent, Krishnan:2020fer,Balasubramanian:2020coy,Balasubramanian:2020xqf,Manu:2020tty,Karlsson:2021vlh,Wang:2021woy,Miao:2021ual,Bachas:2021fqo,May:2021zyu,Kawabata:2021hac,Bhattacharya:2021jrn,Anderson:2021vof,Miyata:2021ncm,Kim:2021gzd,Hollowood:2021nlo,Wang:2021mqq,Aalsma:2021bit,Ghosh:2021axl,Neuenfeld:2021wbl,Geng:2021iyq,Balasubramanian:2021wgd,Uhlemann:2021nhu,Neuenfeld:2021bsb,Kawabata:2021vyo,Chu:2021gdb,Kruthoff:2021vgv,Akal:2021foz,KumarBasak:2021rrx,Lu:2021gmv,Omiya:2021olc,Ahn:2021chg,Balasubramanian:2021xcm,Li:2021dmf,Kames-King:2021etp,Sun:2021dfl,Hollowood:2021wkw,Miyaji:2021lcq,Bhattacharya:2021dnd,Goswami:2021ksw,Chu:2021mvq,Arefeva:2021kfx,Shaghoulian:2021cef,Garcia-Garcia:2021squ,Buoninfante:2021ijy,Yu:2021cgi,Nam:2021bml,He:2021mst,Langhoff:2021uct,Ageev:2021ipd,Pedraza:2021cvx,Iizuka:2021tut, Miyata:2021qsm,Gaberdiel:2021kkp,Uhlemann:2021itz,Collier:2021ngi,Hollowood:2021lsw,emparan2021holographic,omidi2021entropy,bhattacharya2021bath,Yu:2022xlh,Yadav:2022mnv,Lee:2022efh,Liu:2022pan,Afrasiar:2022ebi,Lin:2022qfn,Demulder:2022aij,Bianchi:2022ulu,Lin:2022aqf,Hu:2022ymx,Hu:2022zgy}. 
We found an interesting phase diagram as well as analytical agreement between the field theory and gravity calculations for the more general case where the bipartitions on the two sides of the thermofield double are and are asymmetric, which is a nice check of the AdS/BCFT setup and potentially a valuable result in its own right.

In particular, we found  that for any fixed brane angle, the Page time has strong dependence on the bipartition point only when at least one of the bipartitions is near the defect, or when both are near the event horizon. Physically, this can be understood as a consequence of the system's geometry on the HM surface in each case. In the limit that either bipartition approaches the defect ($u\to 0$), the metric forces the difference between the  areas of the island and the HM surface  to blow up. The opposite limit---the HM surface approaching zero area---can occur only if both bipartitions approach the event horizon simultaneously, as a surface stretching from the event horizon to any point on one of the boundaries still has non-zero area. This is likely saying that the Page time  deviates significantly when the number of degrees of freedom on one or the other side of the bipartition is small. We emphasize that this is true only for the theory on the curved background. For example, if the bulk is a BTZ black hole rather than black string, the boundary is flat and the HM surface has the same area for any symmetric bipartition, in which case the only dependence of the Page time on the bipartition point comes from the island surface. In this case, the significant deviations near the boundary don't apply. We conclude that studying different types of systems, and in particular ones where either or both sides are tractable analytically can lead to further understanding of the evolution of information. In our analysis we show that the asymmetric bipartition leads to different limiting behavior than the symmetric case. The more general results from the asymmetric bipartition give new insights into the field theory interpretation. This is the first exact result of this type.

\section*{Acknowledgement}
We thank Andreas Karch, Severin L\"ust, Rashmish Mishra, Suvrat Raju and David Wakeham for useful discussions and relevant collaborations. HG is very grateful to his parents and recommenders. The work of HG is supported by the grant (272268) from the Moore Foundation ``Fundamental Physics from Astronomy and Cosmology." The work of LR is supported by NSF grants PHY-1620806 and PHY-1915071, the Chau Foundation HS Chau postdoc award, the Kavli Foundation grant ``Kavli Dream Team,'' and the Moore Foundation Award 8342. ES is grateful for the mentorship of HG and LR and for the support of his family and friends. The work of ES is supported by the Harvard College PRISE program.

\bibliographystyle{JHEP}
\bibliography{main}
\end{document}